\documentclass{elsart}
\usepackage{ifpdf}
\usepackage{graphicx,amssymb,lineno}
\usepackage{bm}

\makeatletter
\def\elsartstyle{%
    \def\normalsize{\@setfontsize\normalsize\@xiipt{14.5}}
    \def\small{\@setfontsize\small\@xipt{13.6}}
    \let\footnotesize=\small
    \def\large{\@setfontsize\large\@xivpt{18}}
    \def\Large{\@setfontsize\Large\@xviipt{22}}
    \skip\@mpfootins = 18\p@ \@plus 2\p@
    \normalsize
}
\@ifundefined{square}{}{\let\Box\square}
\makeatother

\begin{document}

\begin{frontmatter}
\title{Statistics of velocity gradients in wall-bounded shear flow turbulence}

\author{Thomas Boeck, Dmitry Krasnov and J\"org Schumacher}
\address{Institute for Thermodynamics and Fluid Mechanics, Technische Universit\"at Ilmenau, P.O.Box 100565, D-98684 Ilmenau, Germany}

\begin{abstract}
The statistical properties of velocity gradients in a wall-bounded turbulent channel flow are 
discussed on the basis of three-dimensional direct numerical simulations. Our analysis is 
concentrated on the trend of the statistical properties of the local enstrophy 
${\bm \omega}^2({\bm x},t)$ and the energy dissipation rate $\epsilon({\bm x},t)$ with increasing distance
from the wall. We detect a sensitive dependence of the largest amplitudes of both fields (which correspond 
with the tail of the distribution) on the spectral resolution. The probability density functions of each 
single field as well as their joint distribution vary significantly with increasing distance from the wall. 
The largest fluctuations of the velocity gradients are found in the logarithmic layer. This is in agreement 
with recent experiments which observe a bursting of hairpin vortex packets into the logarithmic region.  
\end{abstract}

\end{frontmatter}

\section{Introduction}
\label{intro}

Turbulence is associated with large fluctuations of velocity gradients which 
appear preferentially at the smallest scales of the flow. The amplitudes of 
the fluctuations exceed the mean values by orders of magnitude when the Reynolds
number of the flow is sufficiently large. This behaviour is known as 
small-scale intermittency \cite{Sreenivasan1999}. It has been discussed recently that
a deeper understanding of fluid turbulence as a whole requires a detailed resolution 
of the intermittent dynamics at the small-scale end of the inertial range \cite{Sreenivasan2004}.
This will include scales that are smaller than the (mean) Kolmogorov dissipation length
$\eta_K=\nu^{3/4}/\langle\epsilon\rangle^{1/4}$ with the kinematic viscosity $\nu$ and the mean
energy dissipation rate $\langle\epsilon\rangle$. Although significant progress in 
measurement techniques has been made, the finest structures remain still spatially 
unresolved \cite{Zeff2003} or the flows are limited to very low Reynolds numbers 
\cite{Mullin2006}. Direct numerical simulations can reach today sufficiently high Reynolds 
numbers while resolving the gradients at the small-scale end of the inertial range 
cascade properly \cite{Schumacher2007,Donzis2008,Horiuti2008}. 

The small-scale structure and statistics of turbulence has been mostly studied for the case
of homogeneous, isotropic and statistically stationary turbulence. A wall-bounded shear flow consists 
of a boundary layer which is dominated by coherent streamwise structures and a central bulk region 
in which they are basically absent. The statistics in the wall-normal direction is inhomogeneous and 
requires a height-dependent analysis on account of the varying strength of the mean shear. 
As a consequence, there is significantly less work on this aspect for wall-bounded flows than for homogeneous isotropic
turbulence
\cite{Kim1993,Saddoughi1994,Blackburn1996,Jimenez2006}. 
The coupling between large-scale shear and intermittency in the inertial cascade range has been
studied in channel flow simulations \cite{Toschi1999,Toschi2000}. Further studies in the same 
flow configuration demonstrated the violation Kolmogorov's refined similarity hypothesis (K62) \cite{Kolmogorov1962} 
when approaching the walls \cite{Benzi1999,Casciola2001}. It was therefore suggested to replace the
third-order increment moment in  K62 by a second-order one which is directly coupled to the large-scale
shear in the corresponding balance equations. As a consequence, shear-improved eddy viscosities 
in Smagorinsky subgrid-scale models can improve the representation of Reynolds stresses as shown in Ref.
\cite{Leveque2007}.    

In comparison with homogeneous, isotropic and statistically stationary turbulence, 
the presence of an inhomogeneous direction increases
the level of flow complexity significantly. With regard to the 
velocity gradient statistics one can identify the following two aspects.  
First, it is more challenging to measure all nine derivatives of the velocity 
gradient tensor with the sufficient resolution in an open shear-flow-setup (see e.g. \cite{Adrian2007}). 
Second,  it is frequently believed that shear flow turbulence is in a state of local isotropy at 
the small-scale end for larger Reynolds numbers \cite{Saddoughi1994}. More recent 
experimental and numerical studies demonstrated however that significant deviations persist,
in particular when higher-order moments are discussed \cite{Shen2000,Schumacher2003,Biferale2005}. 
It can thus be expected that some fundamental differences to the homogeneous isotropic case persist
to very high Reynolds numbers. All this suggests to our view a systematic study of the height-dependence 
of the statistics of the velocity gradient fields in turbulent shear flows.  
    
In the present work, we want to make a first step in this direction and conduct an analysis 
of the small-scale statistics of the velocity gradient in a wall-bounded shear flow. It is based 
on three-dimensional direct numerical simulations of a turbulent channel flow. Our study will be focussed to two
fields, the local enstrophy field and the energy dissipation rate field. The gradient tensor of
the velocity fluctuations, $m_{ij}$, can be decomposed into a symmetric and an antisymmetric part
\begin{equation}
m_{ij}=\frac{\partial v_i^{\prime}}{\partial x_j}=a_{ij}+s_{ij}\,.
\end{equation}
The local enstrophy field, 
${\bm \omega}^2=({\bm\nabla\times v^{\prime}})^2$, probes the magnitude of the antisymmetric part 
$a_{ij}=-\epsilon_{ijk}\omega_k/2$ of the velocity fluctuation gradient tensor. 
Here $\epsilon_{ijk}$ is the fully antisymmetric Levi-Civita tensor. The energy dissipation rate field, 
$\epsilon=2\nu s_{ij}s_{ji}$, measures basically the magnitude of the symmetric part $s_{ij}$ of 
$m_{ij}$. The ensemble average of both fields are connected by
\begin{equation}
\langle\epsilon\rangle=\nu \langle {\bm \omega}^2\rangle
\label{epsom2}
\end{equation}
for homogeneous isotropic turbulence. This does however not imply that the local statistical fluctuations
are synchronized \cite{Zeff2003,Chen1997}. In addition to homogeneous isotropic turbulence, we have 
to conduct the statistical analysis for both fields as a function of distance from the wall and to separate 
the turbulent fluctuations from the mean flow. 

The outline of the paper is as follows. We discuss the equations of motion and the numerical 
model in the next section. Afterwards we investigate the dependence of the small-scale statistics 
on the spectral resolution which sets the stage for the analysis. The subsequent section 
investigates the single-field and joint statistics of the local enstrophy and dissipation rate 
fields as a function of the distance from the wall. We summarize our findings and give a brief 
outlook at the end. 

\begin{figure}
\begin{center}
\includegraphics[angle=0,scale=0.5,draft=false]{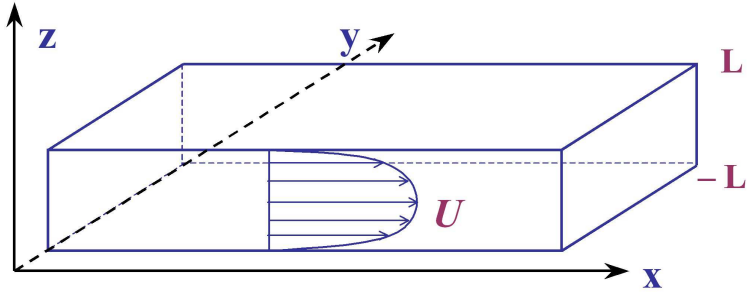}
\caption{Sketch of the channel flow geometry and the coordinates. The simulation 
domain is $L_x \times L_y\times L_z= 4\pi \times 2\pi \times 2$ in units of the
half-width $L$.} 
\label{fig1}
\end{center}
\end{figure}
\section{Equations and numerical model}
The numerical method solves  the Navier-Stokes equations
for an incompressible flow which are given in dimensionless form by
\begin{equation}
\label{nsviscous}
\frac{\partial \bm{v}}{\partial t} -
\bm{v}\times\bm{\Omega}
= -\nabla \left(p+v^2/2\right)
+ \frac{1}{{\rm Re}} \nabla^2 \bm{v},\hskip4mm
\nabla \cdot\bm{v}  =  0,
\end{equation}
with $\bm{\Omega}=\nabla\times\bm{v}$ and boundary conditions
\begin{equation}
\label{bctopviscous}
\bm{v}=0 
\hskip3mm\mbox{at $z=\pm1$}.
\end{equation}
Figure \ref{fig1} illustrates the geometry and the coordinates of the channel flow.
The applied pseudospectral method is based on Fourier series in the horizontal 
directions $x$ and $y$ and a Chebyshev polynomial expansion in the 
vertical $z$--direction \cite{Canuto1977,Gottlieb1977}. The solenoidal velocity field is 
represented in terms of two scalar quantities  $\psi$ and $\varphi$ using the 
poloidal-toroidal decomposition 
\begin{equation}
\bm{v}({\bm x},t)= 
\bm{\nabla} \times \left(\bm{\nabla} \times \bm{e}_z\varphi({\bm x},t) \right) +
   \bm{\nabla} \times \bm{e}_z\psi({\bm x},t). 
\end{equation} 
Equations for $\varphi$ and $\psi$ are derived by taking the curl and 
twice the curl of the momentum equation and projection onto the vertical 
direction. We obtain two equations for the vertical velocity component 
$v_z=-\Delta_h \varphi$ and the vertical vorticity component 
$\Omega_z= -\Delta_h \psi$, where $\Delta_h=\partial_x^2+\partial_y^2$
denotes the horizontal Laplace operator. The quantities $v_z$ and $\Omega_z$ 
determine the velocity field up to a mean flow $U(z,t) \bm{e}_x +V(z,t) \bm{e}_y$.
Equations for $U$ and $V$ are obtained by averaging the momentum
equation over horizontal cross-sections of the periodicity domain.
Upon introducing the definitions 
\begin{equation}
\zeta=\Omega_z,\hskip4mm \eta=\nabla^2 v_z,\hskip4mm \xi=v_z
\end{equation}
for ease of notation, the evolution equations based on the poloidal-toroidal 
representation take the form
\begin{eqnarray}
&&\nabla^2 \zeta - {\rm Re} 
\, \partial_t \zeta   = -{\rm Re} \,\bm{e}_z \cdot\left[  \bm{\nabla} 
\times \left(\bm{v} \times  \bm{\Omega}\right)\right],\\
\label{eta-eq}
&& \nabla^2 \eta - {\rm Re} \, \partial_t \eta= {\rm Re}\left[\partial_z  \bm{\nabla} \cdot 
\left(\bm{v} \times  \bm{\Omega}\right)-
\bm{e}_z \cdot  \nabla^2  \left(\bm{v} \times  \bm{\Omega}\right)\right],\\
\label{xi-eq}
&& \nabla^2 \xi = \eta,\\
\label{meanflow-x-eq}
&& \partial_z^2 U-{\rm Re}\, \partial_t U =
{\rm Re}\,\left[ \partial_z \left\langle v_x v_z\right\rangle_A + \partial_x p \right], \\
&& \partial_z^2 V-{\rm Re}\, \partial_t V=
{\rm Re}\,\left[ \partial_z \left\langle v_y v_z\right\rangle_A  \right]. 
\end{eqnarray}
The angular brackets $\langle\cdot\rangle_A$ denote horizontal averages. The 
mean flow equations also contain the  mean driving pressure gradient  $\partial_x p$.
The appropriate boundary conditions are readily derived using the incompressibility 
constraint $\nabla \cdot\bm{v}=0$. We obtain
\begin{equation}
\xi=\zeta=\partial_z\xi=U= V=0 \hskip1cm \mbox{at} \hskip4mm z=\pm 1\,.
\end{equation}
For the time discretization of the evolution equations we use a 
method of second-order accuracy. If we write such an equation 
symbolically as $\partial_t f = {\cal L} f + {\cal N}(f)$ where $\cal L$ denotes 
a linear operator and $\cal N$ the remaining terms, our time-stepping scheme reads
\begin{equation}
\frac{3 f^{n+1}-4 f^n + f^{n-1}}{2 \Delta t}= {\cal L} f^{n+1} 
+ 2{\cal N}(f^n)-
{\cal N}(f^{n-1})\,,
\end{equation}
where $\Delta t$ is the time step. The left hand side approximates the time 
derivative $\partial_t f$ at the time level $n+1$ using the previous two time 
levels. The linear term ${\cal L}f$ is treated implicitely, and $\cal N$ 
explicitly through the second-order Adams-Bashforth method, where the prefactors 
correspond to a linear extrapolation to the time level $n+1$ \cite{Krasnov2004}. 
The solution procedure for the coupled eqns. (\ref{eta-eq}) and (\ref{xi-eq}) requires auxiliary
functions to satisfy the boundary conditions $\partial_z \xi=0$ for eq. (\ref{eta-eq})
as explained in \cite{Krasnov2004}. The numerical code implements distributed-memory
parallelization through MPI.

\begin{figure}
\begin{center}
\includegraphics[angle=0,scale=0.5,draft=false]{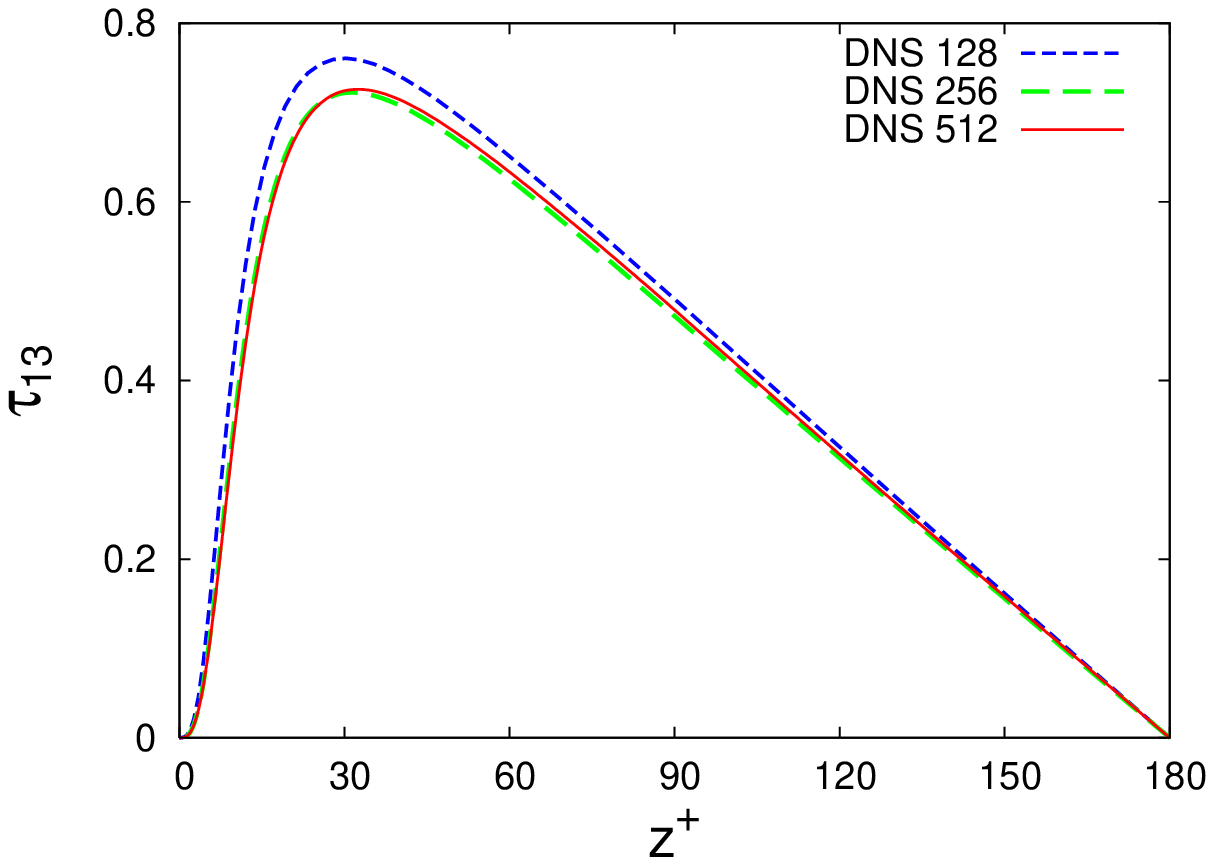}
\includegraphics[angle=0,scale=0.5,draft=false]{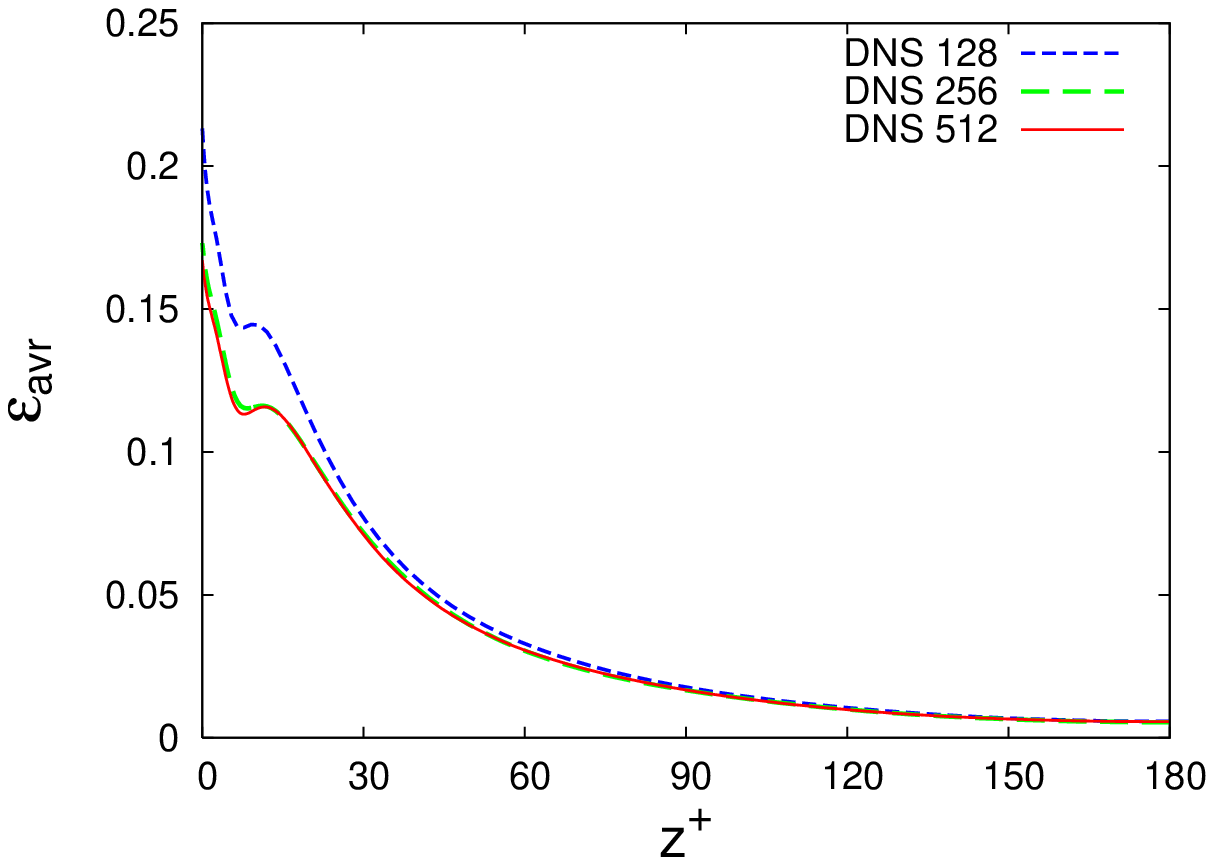}
\caption{Vertical profiles of the Reynolds shear stress $\tau_{13}$ (left) 
and the energy dissipation rate $\epsilon$ (right) averaged in planes of constant $z$.
The wall-normal coordinate $z$ is given in wall units. The results of 
small- ($128^3$ points, blue curves), medium- ($256^3$ points, green curves) and 
high-resolution ($512^3$ points, red curves) simulations are shown. In each case,
the quantities are made nondimensional with 
the wall friction velocity $u_\tau$ from the simulation with $512^3$ points. }
\label{fig2}
\end{center}
\end{figure}
\section{Single quantity statistics}
\subsection{Spectral resolution effects}
Our statistical analysis is focussed to the small-scale properties of the 
channel flow turbulence. In order to set the stage for this investigation, 
we first compare  statistical properties at three different spectral resolutions and the 
same set of parameters. The first case with $128^3$ (or more precisely $128^8\times 129$) grid 
points corresponds with the classical simulation\footnote{More precisely, the grid resolution in 
\cite{Kim1987} was $N_x\times N_y\times N_z=190\times 129\times 160$ (with $y$ being the 
wall-normal coordinate) for a channel with the same sidelengths as in the present work.} 
of Kim, Moin and Moser \cite{Kim1987} in terms of the vertical resolution. The second case is 
two times better resolved in each space direction with $256^3$ grid points. The highest spectral 
resolution is  four times better resolved than the first  one and contains $512^3$ (or more precisely 
$512^2\times 513$) collocation points. The three runs  were performed with identical fixed volume flux
such that the resulting friction Reynolds number approaches the grid-independent
limit  $Re_{\tau}=180$.  The definition
\begin{equation}
Re_{\tau}=\frac{u_{\tau}L}{\nu}\,.
\end{equation}
is based on the wall friction velocity $u_{\tau}$, the half-channel height  
$L$, and the kinematic viscosity by $\nu$. We follow the usual convention and 
denote the turbulent velocity fluctuations by 
\begin{equation}
{\bm v}^{\prime}({\bm x},t)={\bm v}({\bm x},t)-U(z){\bm e}_x\,.
\end{equation}
Figure \ref{fig2} (left) compares the Reynolds shear stress $\tau_{13}=\langle 
v_x^{\prime} v_z^{\prime}\rangle_{A,t}$ as a function of $z^+=z \nu/u_{\tau}$. 
The three vertical profiles of $\tau_{13}$ coincide quite well, 
except for slight deviations around the maximum in case of the lowest resolution. 
The additional subscript $t$ in $\langle \cdot\rangle_{A,t}$ indicates time averaging.
We repeat the analysis for the vertical profile of the
energy dissipation rate averaged 
over time and in planes at constant distance from the wall. The quantity is defined as
\begin{equation} 
\epsilon_{avr}(z)=\langle\epsilon({\bm x},t)\rangle_{A,t}=2\nu\langle s_{ij}s_{ji}\rangle_{A,t}=\frac{\nu}{2} 
\left\langle\left(\frac{\partial v_i^{\prime}}{\partial x_j}+\frac{\partial v_j^{\prime}}{\partial x_i}
\right)^2\right\rangle_{A,t}\,.
\end{equation}
The plot yields almost perfectly collapsing curves for most of the channel. Deviations for the
smallest resolution  exist for the viscous layer at $z^+\le 20$. We can conclude 
that the lowest resolution is sufficient when first-order moments are considered, 
such as shear stresses. It is therefore suitable for many purposes, such as validations 
of closures of the Reynolds averaged equations \cite{Wilcox}. We can however see already, 
that the mean dissipation rate profile deviates significantly in the viscous layer.  

A full statistical information requires the knowledge of all moments or the
probability density function (PDF).  We proceed therefore by comparing the PDFs of the 
energy dissipation rate, $\epsilon({\bm x},t)$, and the local enstrophy, 
${\bm \omega}^2({\bm x},t)=({\bm\nabla\times v^{\prime}})^2$, for both resolutions.  
\begin{figure}
\begin{center}
\includegraphics[angle=0,scale=0.53,draft=false]{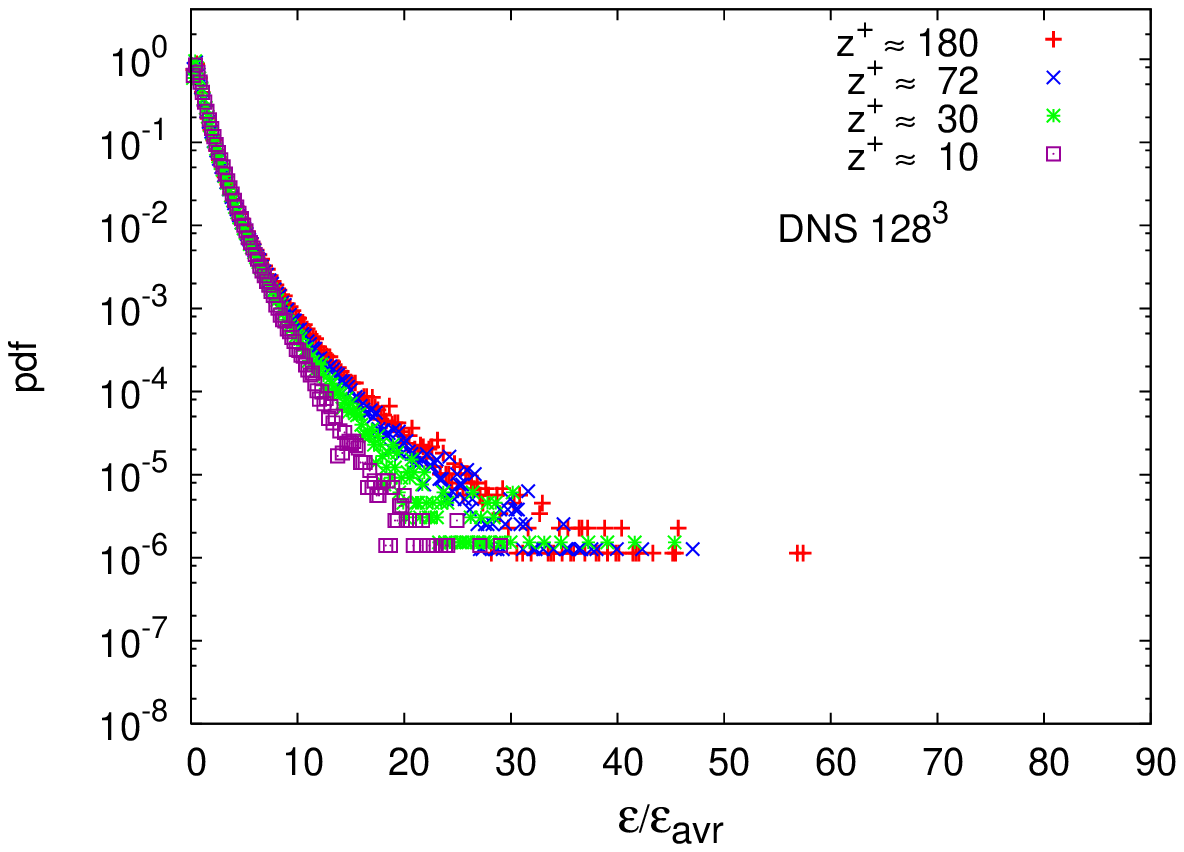}
\includegraphics[angle=0,scale=0.53,draft=false]{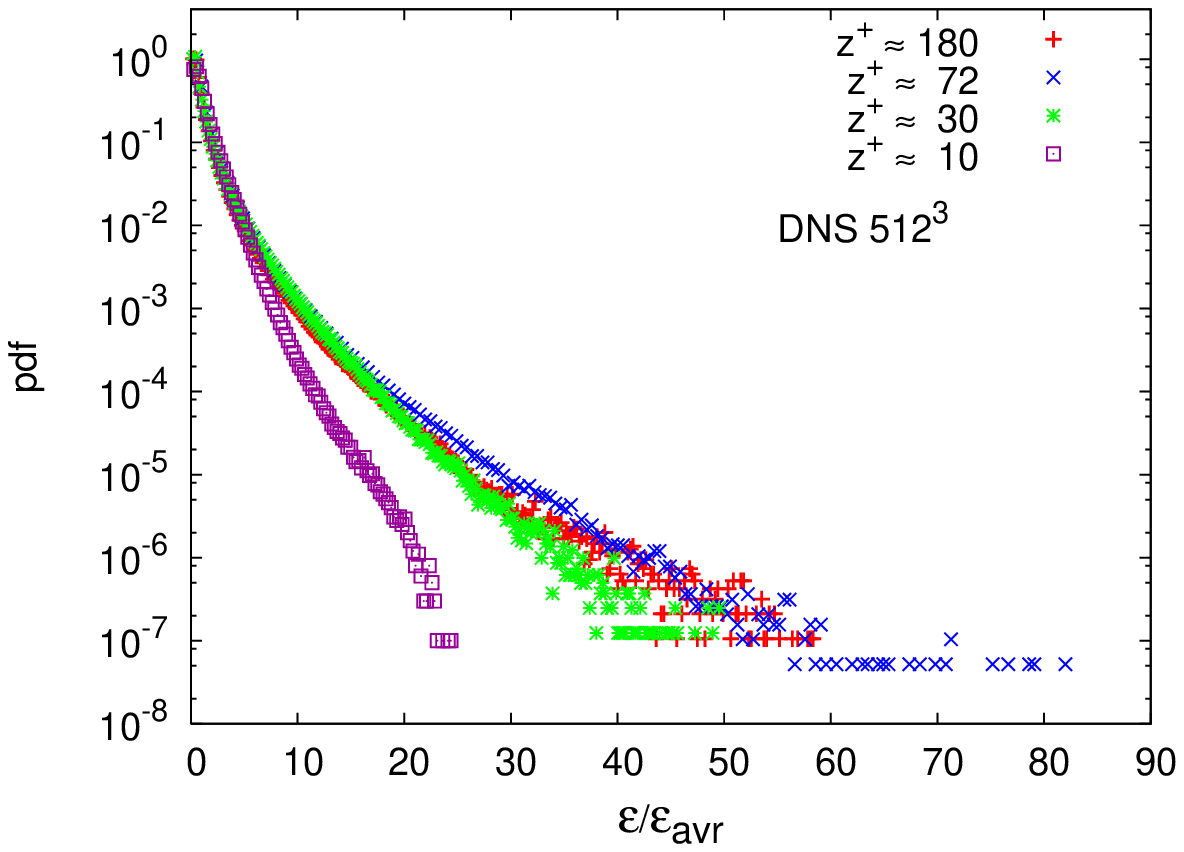}
\includegraphics[angle=0,scale=0.53,draft=false]{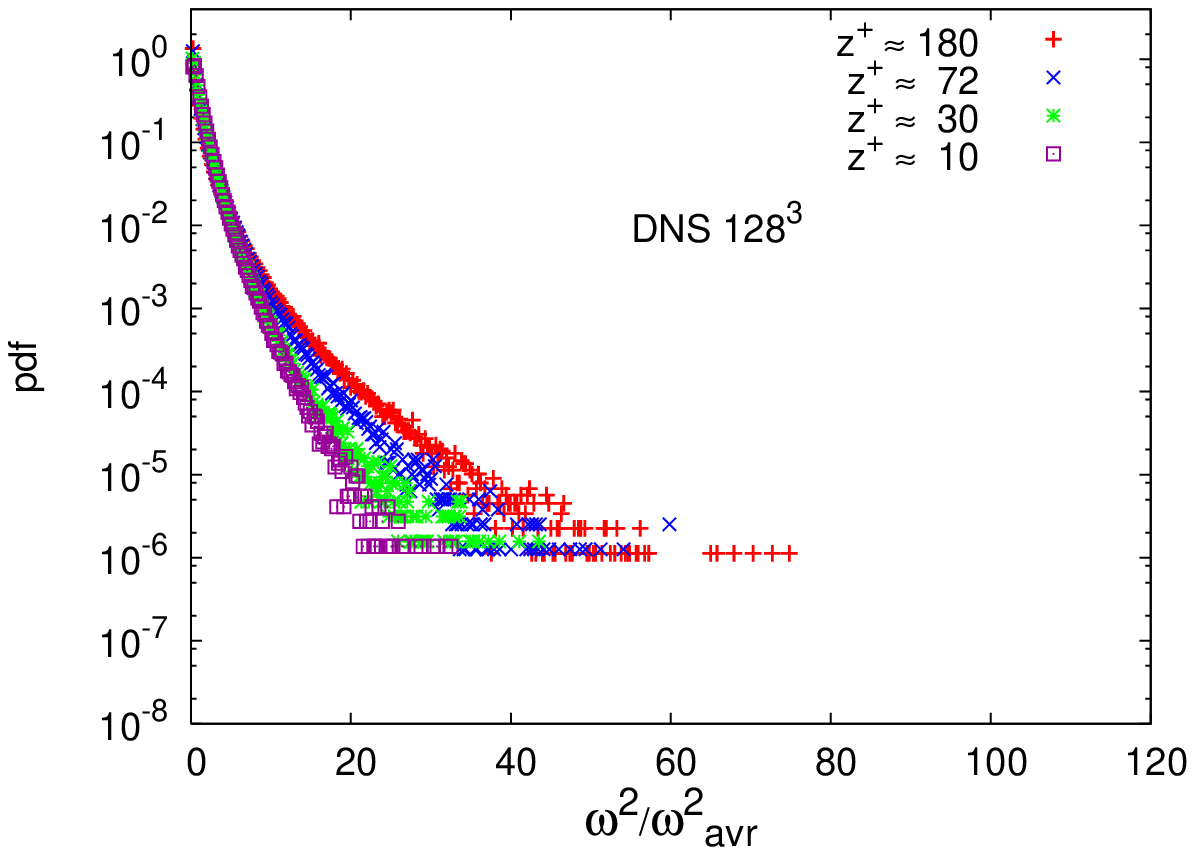}
\includegraphics[angle=0,scale=0.53,draft=false]{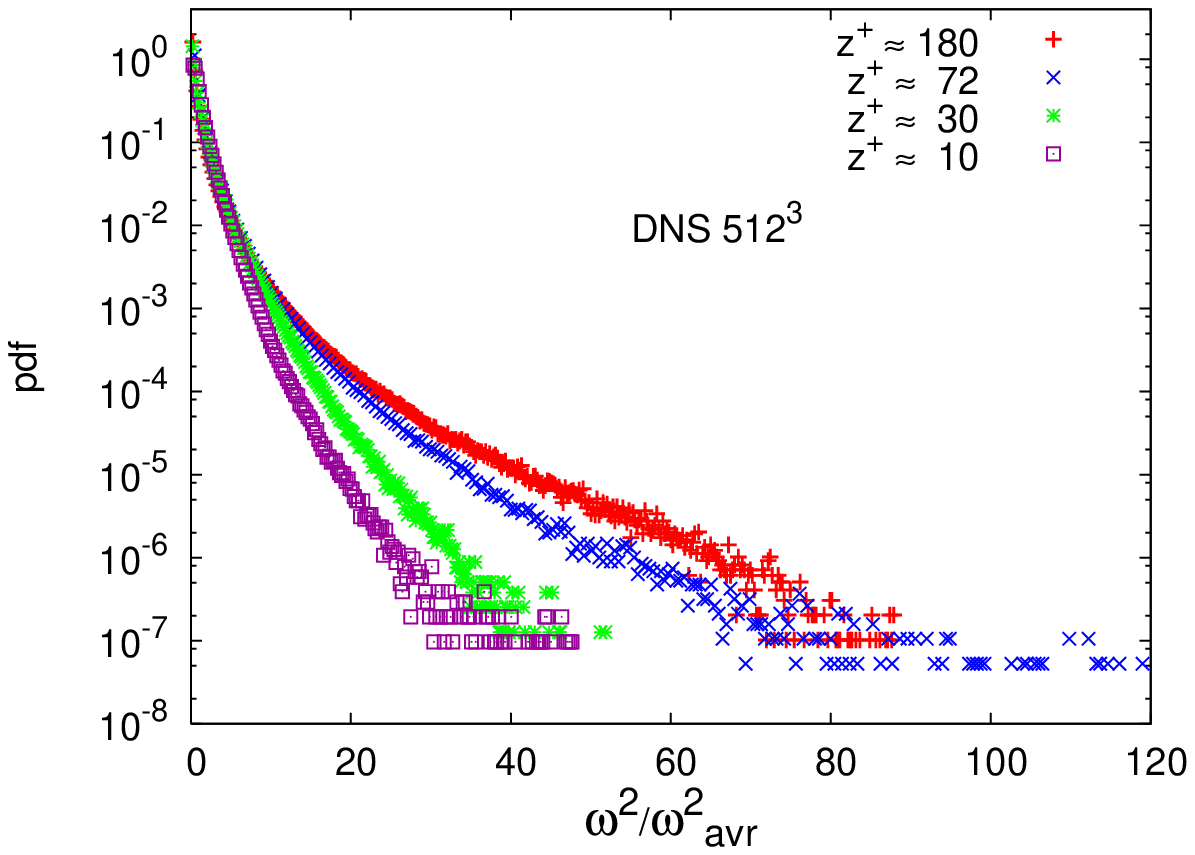}
\caption{Comparison of the probability density functions of the energy dissipation 
rate (upper row) and local enstrophy (lower row) for the two spectral resolutions. 
Data are obtained in planes at four different distances from the wall which are indicated 
in the legends. The left column stands for the lowest resolution case, the right one for the
highest resolution. Both quantities are given in units of their statistical mean in the 
corresponding plane. We denote $\epsilon_{avr}=\langle\epsilon\rangle_{A,t}$ and 
$\omega^2_{avr}=\langle\omega^2\rangle_{A,t}$.}
\label{fig3}
\end{center}
\end{figure}
Four different distances have been chosen: $z^+=10$ is in the viscous buffer layer,
$z^+=30$ at the begining of the loga\-rithmic layer where the Reynolds shear stress
has its maximum (see Fig. \ref{fig2} (left)), $z^+=72$ in the loga\-rithmic region, 
and $z^+=180$ in the midplane of the channel. While the first-order moments are
basically unchanged at the higher resolution, we can detect clear differences 
in the far tails of the PDFs for all cases. The far tails are exactly the parts of the PDF
that become relevant for higher-order moments. Their sufficient resolution requires
the efforts made in the present work (see also \cite{Schumacher2007} for a detailed discussion). 

\begin{table}[h]
\caption{Moments of the normalized energy dissipation rate as a function of height and resolution. }

\begin{center}
\begin{tabular}{@{}|c|ccc|ccc|ccc|@{}}
\hline
$~~~$ & \multicolumn{3}{|c|}{$<\epsilon^2>$} & \multicolumn{3}{|c|}{$<\epsilon^3>$} & \multicolumn{3}{|c|}{$<\epsilon^4> / <\epsilon^2>^2$}\\
\hline
$z^+$ & $128$ & $256$ & $512$ & $128$ & $256$ & $512$ & $128$ & $256$ & $512$\\
\hline
$ 10$ & $2.12$ & $2.15$ & $2.14$ & $ 7.73$ & $ 7.93$ & $ 7.83$ & $ 9.59$ & $ 9.45$ & $ 9.34$\\
$ 30$ & $2.25$ & $2.93$ & $3.14$ & $ 9.92$ & $17.95$ & $21.83$ & $15.45$ & $20.40$ & $25.44$\\
$ 72$ & $2.46$ & $2.96$ & $3.15$ & $13.00$ & $20.43$ & $24.36$ & $21.34$ & $32.53$ & $36.77$\\
$180$ & $2.55$ & $2.81$ & $3.06$ & $14.56$ & $18.67$ & $23.07$ & $25.24$ & $30.47$ & $35.94$\\
\hline
\end{tabular}

\end{center}
\label{tab:epsilon}
\end{table}

\subsection{Height dependence of statistics}
Let us turn now to the high-resolution case of both fields in the right column 
of Fig. \ref{fig3}. Note first that the amplitudes are always rescaled by the corresponding mean 
values for the given plane at fixed $z^+$. Close to the wall, the turbulent velocity fluctuations are small. 
The mean dissipation profile indicates however that this is not the case for the magnitude of the 
derivatives. As we can see in Fig. \ref{fig2}, the mean energy dissipation rate is largest
close to the wall and drops by more than two orders of magnitude toward the channel centre plane. 
The tails of both distributions are the sparsest ones in the vicinity of the wall. While the means are large 
the variations about the means remain small. The extension of the tails varies with increasing 
distance from the wall as the data for the three other values of $z^+$ indicate. Since the mean values 
decrease monotonically towards the midplane, high-amplitude events are shifted ever further into the tail. 
The fattest tails are observed at $z^+=72$, i.e. in the middle of the logarithmic region. 
Particularly here, high amplitudes can be expected further away from the wall
due to bursting events. They are for example caused by coherent hairpin vortices that can grow 
out into the logarithmic layer. Such a large-scale hairpin vortex packet formation has been recently 
suggested by Adrian and co-workers on the basis of experiments and direct numerical simulations 
\cite{Adrian2007}. Our statistical analysis is consistent with this observation.

The high resolution case reproduces another feature which is well-known from the homogeneous
isotropic case. The tail of the local enstrophy is fatter than that of the energy 
dissipation rate \cite{Zeff2003,Donzis2008}. Simulations of isotropic box turbulence by Donzis 
{\it et al.} \cite{Donzis2008} suggest that this difference in the tails decreases with increasing
Reynolds number, a point that certainly has to be studied in the shear flow case as well.

Height and resolution dependences of higher moments of the dissipation rate
are given in table \ref{tab:epsilon}. The fatter tails of the PDFs at higher resolution account
for the growth in the second and third moment with the resolution
for all $z^+$ values except in the viscous region at $z^+=10$. Likewise, the flatness 
$\langle \epsilon^4\rangle/\langle \epsilon^2\rangle^2$ is growing with the
resolution. This quantity is also indicative of the degree of 
intermittency at the smallest lengthscales.  
Remarkably, it is not constant, but increases significantly from
$z^+=10$ to $z^+=72$, i.e. the level of small-scale intermittency depends
on the wall distance outside the viscous region.
 However, it changes only slightly from $z^+=72$ to the midplane.

Regarding the relation between planar averages of enstrophy and energy
dissipation we remark that they are 
connected by the relation
\begin{equation}
\langle \epsilon\rangle_A=\nu \langle{\bm\omega}^2\rangle_A +\nu \partial_z^2\langle
v_z^{\prime\, 2}\rangle_A\, 
\label{epsom3}
\end{equation}
in a wall-bounded shear flow.
In contrast to Eq. (\ref{epsom2}), this relation contains  an additional derivative term
arising from the averaging which is restricted to planes of fixed $z$. We found that the second term on the right 
hand side of (\ref{epsom3}) is at least two orders of magnitude smaller than $\langle{\bm\omega}^2\rangle_A$.
We have also verified that relation (\ref{epsom3}) is satisfied in our numerical simulations. 
Furthermore, the statistics of the energy dissipation rate field deviated from the log-normal distribution
for all four planes, particularly in the far tails at the small and large amplitudes. This property is similar to 
isotropic turbulence.   

\section{Joint statistics of energy dissipation and local enstrophy}  
\begin{figure}
\begin{center}
\includegraphics[angle=0,scale=0.38,draft=false,clip=]{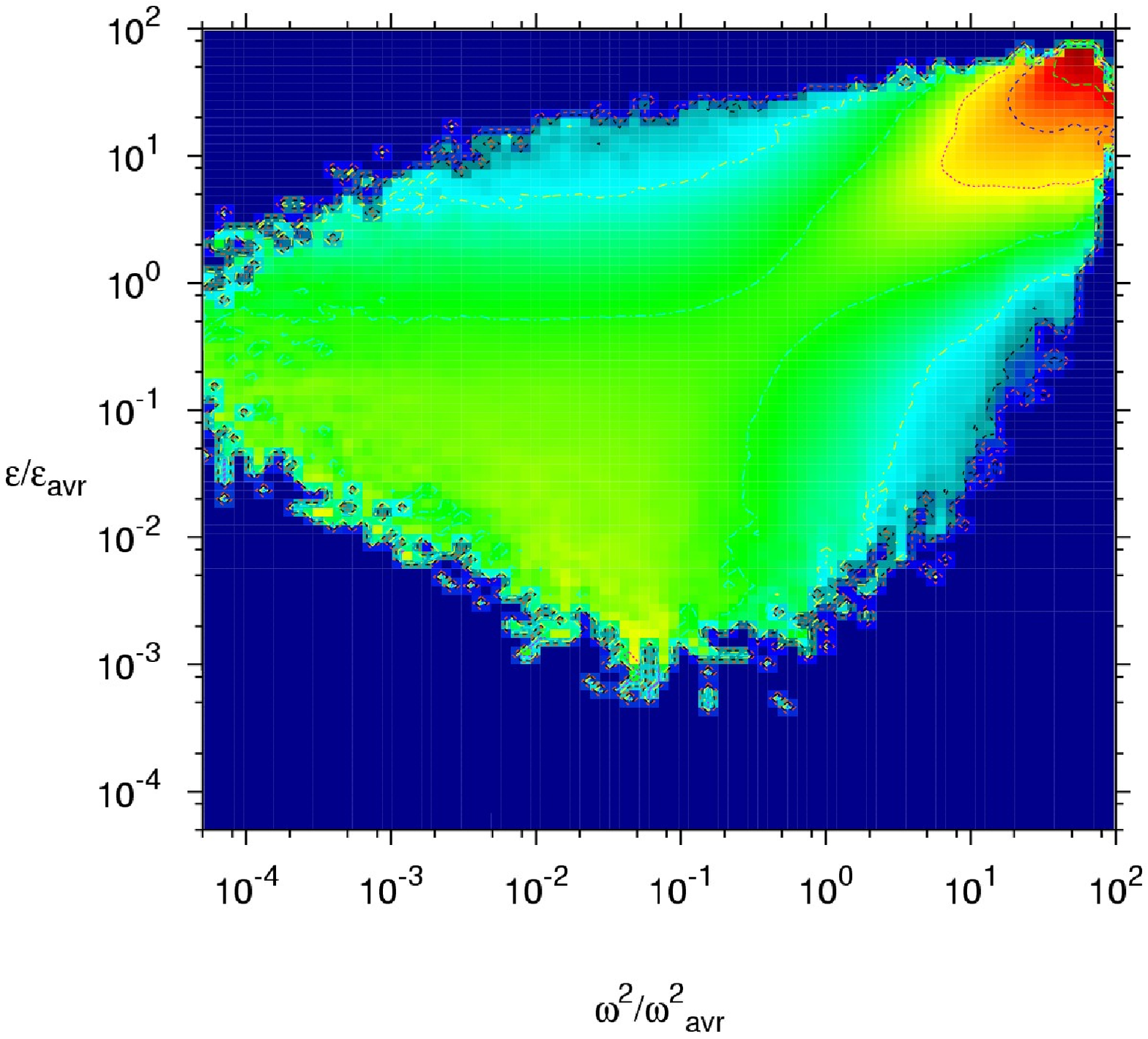}
\includegraphics[angle=0,scale=0.38,draft=false,clip=]{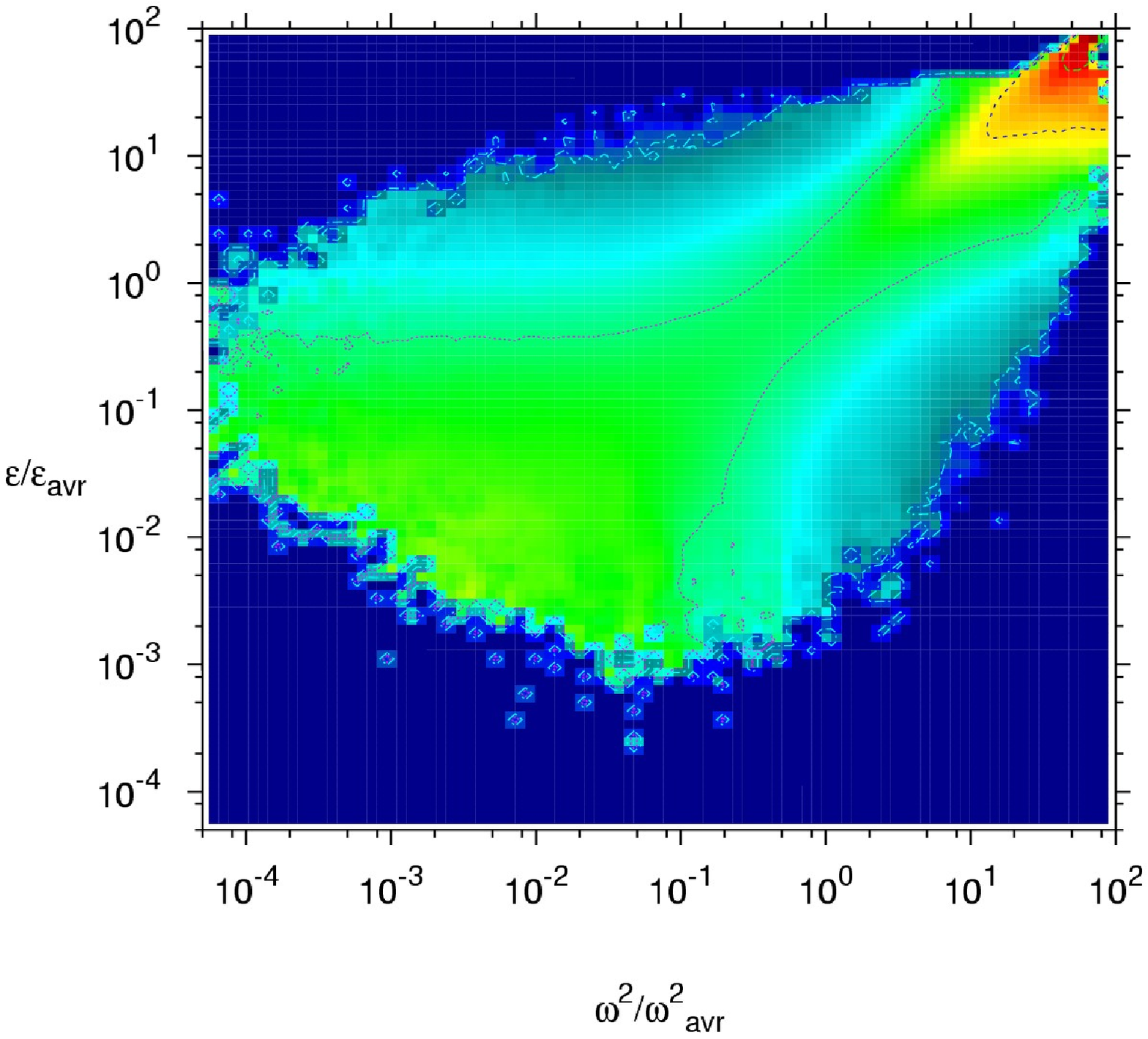}
\includegraphics[angle=0,scale=0.38,draft=false,clip=]{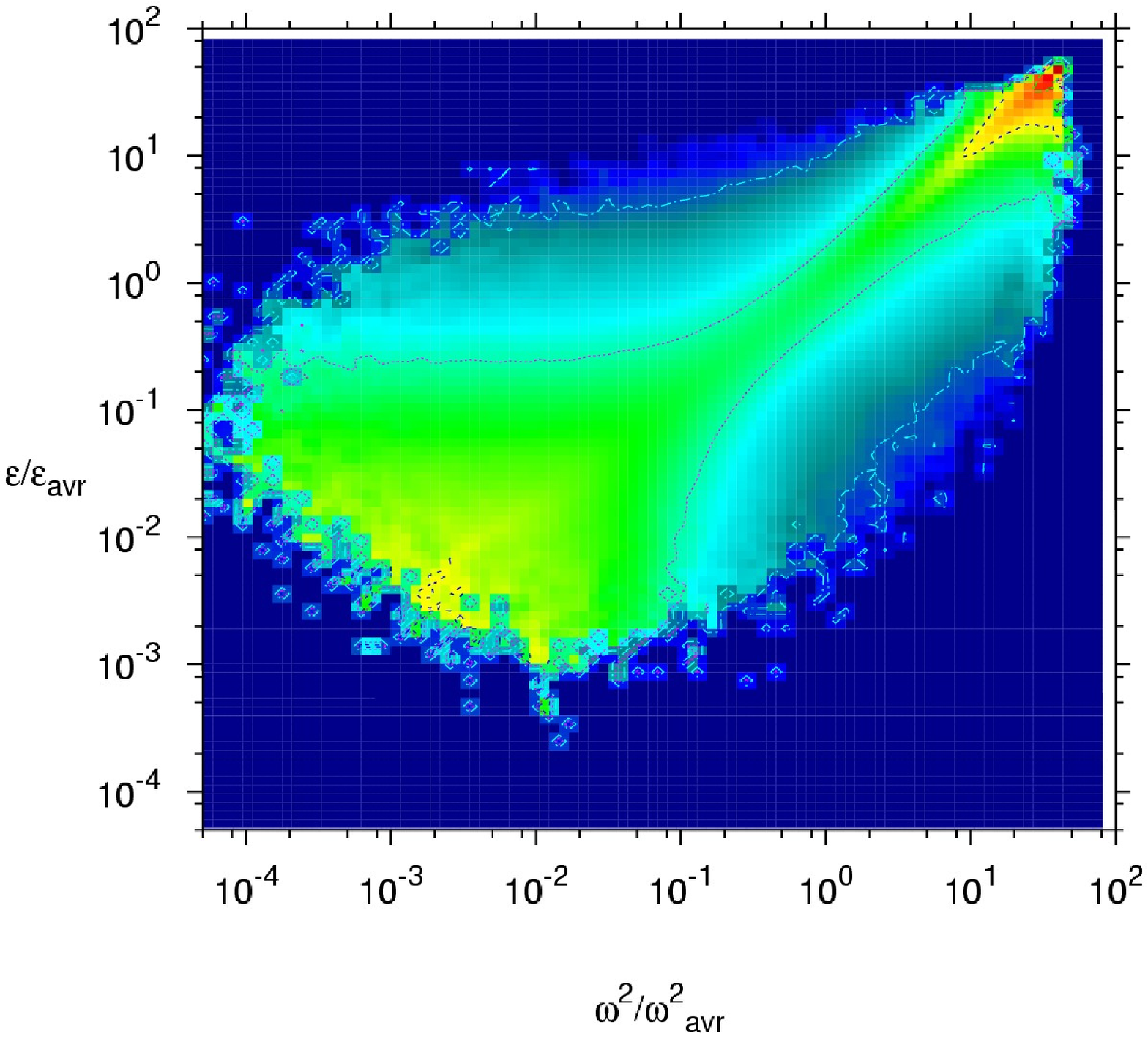}
\includegraphics[angle=0,scale=0.38,draft=false,clip=]{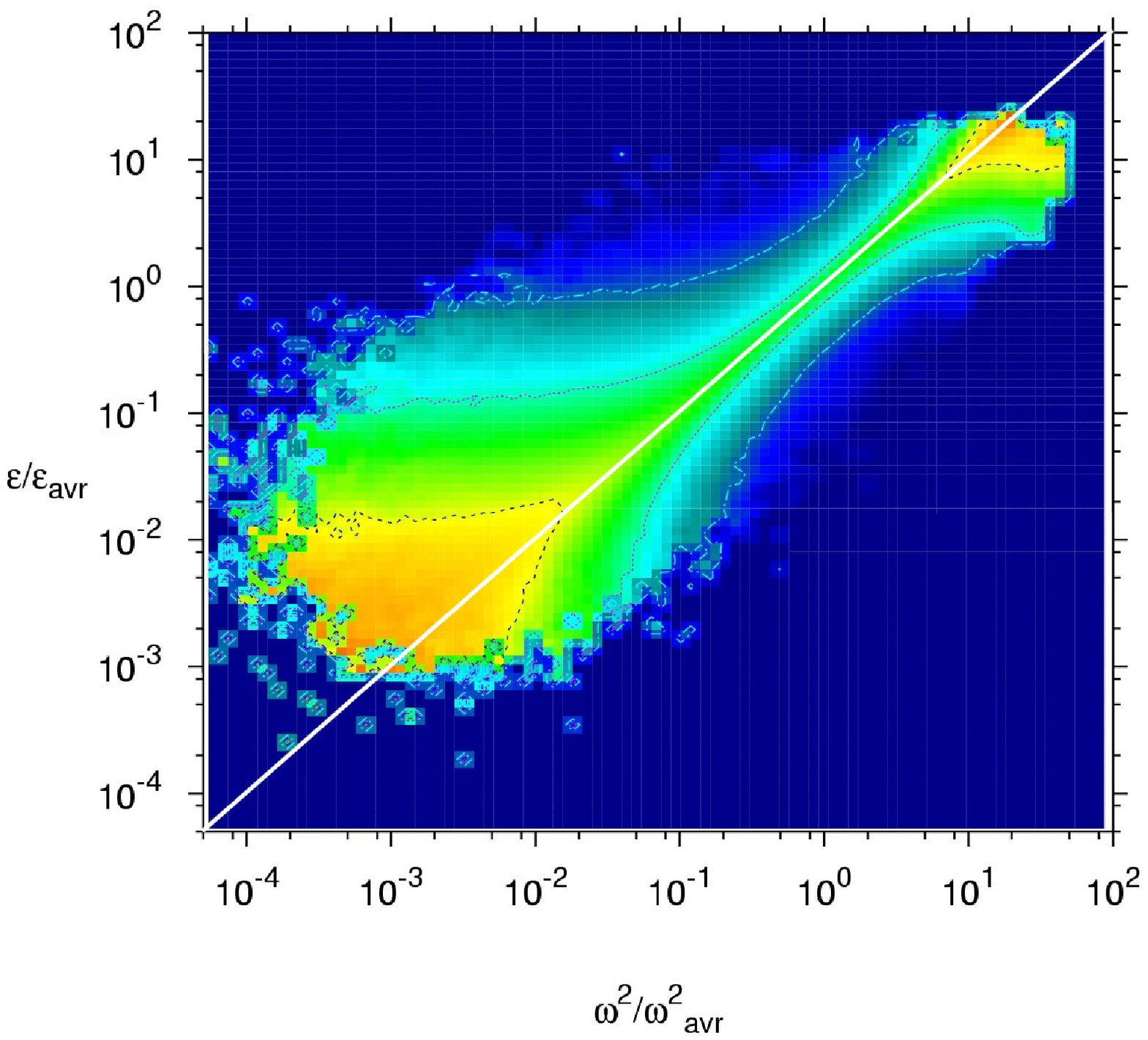}
\caption{Joint and normalized PDFs for the energy dissipation rate
$\epsilon$ and the square of vorticity $\omega^2$ as given by Eq. (\ref{jointpdf}).
Data are taken again at four $z^+$-positions. Upper left: $z^+ = 180$. Upper right:  
$z^+ = 72$. Lower left: $z^+ = 30$. Lower right: $z^+ = 10$. Results of DNS with
$512^3$ grid points are shown, the joint PDF fields are normalized by the product of
separate PDFs for $\epsilon / \epsilon_{avr}$ and $\omega^2 / \omega^2_{avr}$. The 
white diagonal line in the lower right figure corresponds with $\epsilon=\nu{\bm\omega}^2$.
The colour coding is the same in all four figures and given in units of decadic logarithm.
It varies from $10^0$ (red) to $10^{-4}$ (blue).}
\label{fig4}
\end{center}
\end{figure}

In view of the differences in the PDFs for enstrophy and energy dissipation it is
interesting  to study how both fields are statistically correlated.
For this reason we plot in Fig. \ref{fig4} the joint PDF
of $\epsilon$ and ${\bm \omega}^2$ normalized by both single quantity PDFs
\begin{equation}
\Pi(\epsilon, {\bm \omega}^2)=\frac{p(\epsilon, {\bm \omega}^2)}{p(\epsilon)\,p({\bm \omega}^2)}
\label{jointpdf}
\end{equation}  
High amplitudes of $\Pi$ indicate a strong statistical correlation between both fields.
The PDF of the data in the midplane has the broadest support. Very high-amplitude events of 
dissipation and vorticity are strongly correlated as indicated by the local maximum which is 
stretched out into the upper right corner of the plane. This maximum is even more pronounced in the 
logarithmic layer as can be seen in the upper right panel of Fig. \ref{fig4}. The local 
maximum of $\Pi$ in the outer right corner becomes narrower with decreasing distance from the wall. 
It is in line with an overall decrease of the support for highly correlated events. The observed 
trend is consistent with our observations for the single quantity statistics. Again, we can detect
a decrease of the fluctations of both fields about the mean. As a guide to the eye, we added the line 
$\epsilon=\nu {\bm\omega}^2$ (in white) to the lower left panel, i.e. to the joint-PDF plot closest 
to the wall. It can be seen that  a broad range of amplitudes of both fields is basically concentrated now
around this line, which connects the means of both fields (see eqns. (\ref{epsom2}) or 
(\ref{epsom3})). To conclude, at all 
distances from the wall there is a strong statistical correlation between the local enstrophy and the 
dissipation. Close to the wall high-amplitude events of both fields seem to become synchronized.
The support of the statistically correlated events varies significantly across the channel.
 
\begin{figure}
\begin{center}
\includegraphics[angle=0,scale=0.65,draft=false]{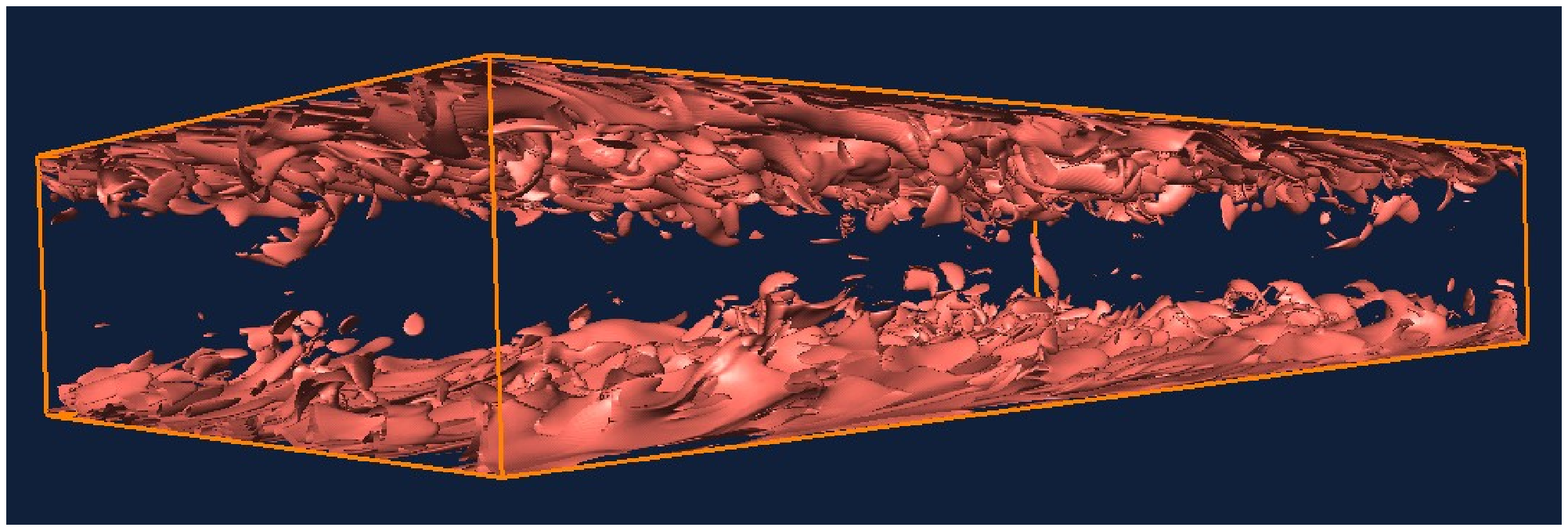}
\vspace{0.1cm}
\includegraphics[angle=0,scale=0.65,draft=false]{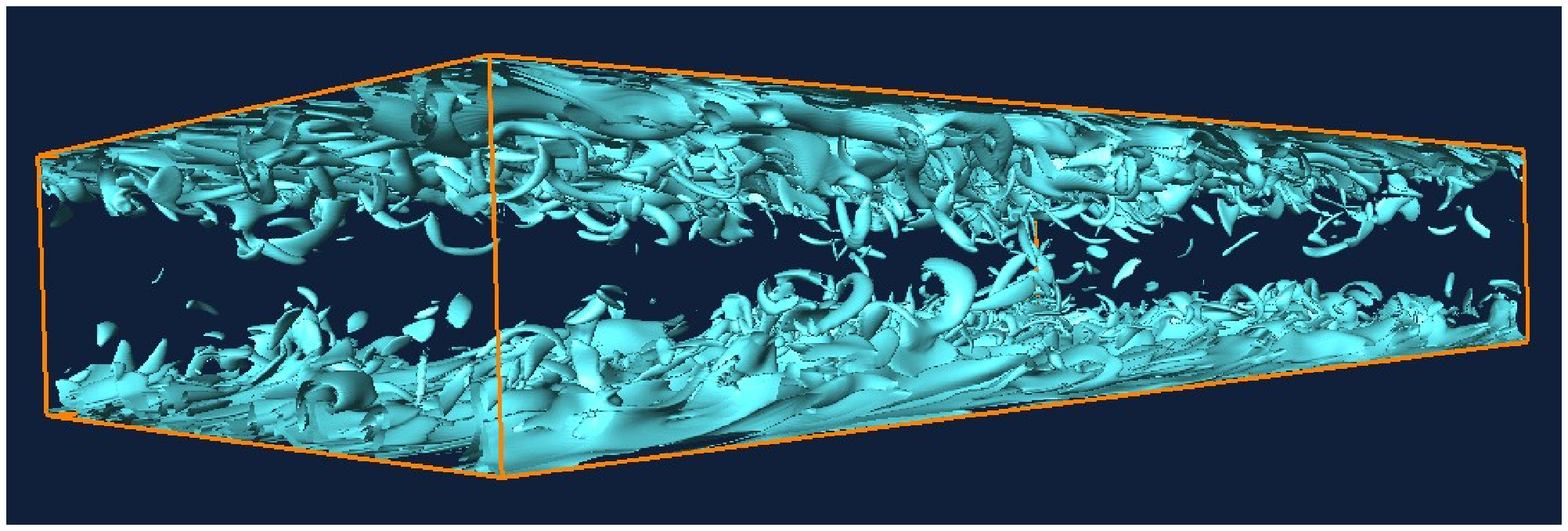}
\caption{Isosurfaces of the energy dissipation rate (upper figure) and the local enstrophy
(lower figure). The isolevels are taken at three times the mean dissipation and the mean 
squared vorticity. }
\label{fig5}
\end{center}
\end{figure}
 
Figure \ref{fig5} displays the structures of both fields. The largest amplitudes
of both fields are clearly concentrated close to the walls.  We also observe the hairpin-like
vortices that burst into the channel center (see lower panel). Isosurfaces of  the energy 
dissipation rate display more sheet-like structures (see upper panel). They reach also out into the logarithmic 
region. Such shear layers can be formed when pairs of streamwise streaks are lifted up 
\cite{Schoppa2000} or when they become linearly unstable. The inspection of the structures 
in both plots supports thus our statistical results obtained before. Maxima of both fields are 
found close together in space.  In particular, we expect, that exactly the bursting events will
contribute significantly to the local maximum of the joint PDF in the upper right panel of Fig. \ref{fig4}.  
A firm proof would require a separation of these coherent structures in the data record, which
will be part of our future work on this subject. 
    
\section{Summary and conclusions}
We have presented a study of the statistics of the energy dissipation rate and 
local enstrophy in a turbulent channel flow. We showed that a full statistical 
analysis of both fields requires a spectral resolution which is larger than the standard 
one. This was one objective of our study, namely to make this case in a turbulent wall-bounded
shear flow. It clearly limits the range of accessible Reynolds numbers. Based on this result, 
we investigated the height-dependence of the statistics of both fields in a first example. 
We showed that the variations of both fields about their mean values are strongest in the 
logarithmic layer. They decrease slightly towards the channel centre and significantly towards
the wall. The 
statistics of the velocity gradients
is thus strongly height-dependent. It is clear that a more comprehensive parametric study has to 
follow, e.g., in order to analyze these trends at different Reynolds numbers. 

There is a further interesting aspect onto which our studies can shed new light.
Recently, new ideas for multifractal subgrid-scale models have been developed which are based on exactly 
such statistical fluctuations of velocity derivatives which have been studied 
here \cite{Burton2005}. The model of \cite{Burton2005} has been successfully validated in homogeneous 
isotropic turbulence and it seems to be a hopeful candidate for  application to shear flow turbulence. 
Investigations in this direction are under way.    
  
{\em Acknowledgements.} Some parts of this work grew out of discussions with 
K. R. Sreenivasan and V. Yakhot. We wish to thank both for their suggestions. 
Our work is supported by the Emmy-Noether-Program (DK and TB) and the 
Heisenberg-Program (JS) of the Deutsche Forschungsgemeinschaft (DFG). The 
supercomputing ressources were provided by the J\"ulich Supercomputing Centre (Germany) under grant HIL01.

\end{document}